\documentclass[runningheads]{llncs}

\usepackage[T1]{fontenc}
\usepackage{amsmath,amssymb}
\usepackage{xcolor}
\usepackage[color]{zed}
\newcommand{\clz}{\color{ZedColor}}
\usepackage{booktabs}
\usepackage{array,tabularx}
\usepackage{listings}
\usepackage{tikz}
\usetikzlibrary{arrows.meta, positioning, shapes.geometric}
\usepackage{microtype}
\usepackage{url}
\usepackage[hidelinks]{hyperref}
\hypersetup{
    colorlinks=true,
    linkcolor=red,
    citecolor=blue,
    urlcolor=cyan,
    pdftitle={CAPRI: Contract-Aware Proof Repair for Isabelle},
    pdfauthor={Jim Woodcock; Gabriel Leite; Augusto Sampaio; Ran Wei}
}

\lstdefinestyle{isabelle}{
  basicstyle=\ttfamily\scriptsize,
  columns=fullflexible,
  keepspaces=true,
  breaklines=true,
  frame=single,
  xleftmargin=0.3em,
  xrightmargin=0.3em
}
\newcommand{\Build}{\mathsf{Build}}
\newcommand{\Conforms}{\mathsf{Conforms}}
\newcommand{\Accept}{\mathsf{Accept}}
\newcommand{\VS}{\textsf{valid-success}}
\newcommand{\FS}{\textsf{false-success}}
\newcommand{\SF}{\textsf{safe-failure}}
\newcommand{\RV}{\textsf{rejected-violation}}

\newcommand{\negbigskip}{\vspace{-\bigskipamount}}

\title{CAPRI: Contract-Aware Proof Repair for Isabelle}
\titlerunning{CAPRI: Contract-Aware Proof Repair for Isabelle}

\author{
  Jim Woodcock\inst{1}
  \and
  Gabriel Leite\inst{2}
  \and
  Augusto Sampaio\inst{2}
  \and
  Ran Wei\inst{3}
}
\authorrunning{J. Woodcock et al.}
\institute{
  Southwest University, China;
  Aarhus University, Denmark;
  University of York, UK \\
  \email{jim.woodcock@york.ac.uk}
  \and
  Centro de Informática, Universidade Federal de Pernambuco (UFPE), Recife, Brazil \\
  \email{gnl2@cin.ufpe.br,acas@cin.ufpe.br}
  \and
  University of Lancaster, UK \\
  \email{r.wei5@lancaster.ac.uk}
}

\begin{document}

\maketitle

\begin{abstract}
We address the use of large language models (LLMs) to help discover Isabelle proofs. An Isabelle build establishes that the submitted theory is accepted, but not that an LLM changed only what the developer authorised. We present CAPRI, a contract-aware repair workflow in which Isabelle checks the proof and an independent checker enforces a machine-readable edit contract. Prompts, proposals, candidate repositories, diagnostics, verdicts, and hashes are retained for audit. We evaluate five workflows on twelve failed proofs from four developments, with three replicates per task and condition, giving 180 runs and 138 valid repairs. Of 144 terminal candidates accepted by Isabelle, six had modified protected text; all arose in iterative workflows that could edit a complete theory. A proof-body-only interface produced 29/36 valid repairs and no contract violations, compared with 31/36 for the corresponding full-theory workflow. One-shot repair produced 22/36, while a later prospectively frozen iterative workflow produced 32/36; these figures compare complete workflows rather than individual mechanisms. A separate post hoc OpenRouter campaign found no improvement in the designated Luna comparisons. A Sol configuration with matched demonstrations produced 33/36 repairs, compared with 29/36 in the frozen OpenAI Responses condition, but the difference was not statistically significant in a one-sided exact McNemar test ($p=0.0625$).

\end{abstract}

\section{Introduction}

Large language models (LLMs) are increasingly used in combination with proof assistants. The basic loop is straightforward: the model proposes a proof, the prover checks it, and any diagnostic may be returned to the model for another attempt. This pattern supports whole-proof generation, tactic search, and recent systems for Isabelle and Lean~\cite{FirstEtAl2023Baldur,ThakurEtAl2023COPRA,Hou2026Isabellm,RequenaEtAl2026Minimal}. Within this loop, Isabelle/HOL provides a strong guarantee: it determines whether the submitted theory is accepted by the prover and its toolchain~\cite{NipkowPaulsonWenzel2002}. It does not, however, determine whether an automated repair stayed within the changes authorised by the developer.

Consider a model asked to repair a single proof so that an Isabelle session builds. Instead of repairing the proof, it might weaken the theorem, add the desired conclusion as an assumption, change a definition or import, or delete a neighbouring regression obligation. It might also introduce commands such as \texttt{sorry}, which admits an unproved theorem, or \texttt{oops}, which abandons a proof attempt. Isabelle may correctly accept the resulting development because it checks the theory it has been given. The failure is not one of prover soundness, but of repair authority: the patch does more than the developer authorised. We call this situation a \emph{false success}. The build succeeds, but the requested repair has not been performed within its assigned boundary.

Our workflow addresses this problem by treating the LLM as an untrusted source of patches. Isabelle checks proof acceptance, while an independent contract checker checks change authorisation. A machine-readable repair contract identifies the protected text and the proof region that may be edited. The checker compares the original and candidate repositories to determine whether the proposed patch conforms to that contract. A candidate is promoted only when it passes both the Isabelle build and the contract check. The accompanying audit record preserves the original repository, contract, prompts, model proposals, candidate trees, contract reports, prover output, and run history. The two acceptance decisions can therefore be inspected and reproduced independently.

The evaluation addresses three research questions:
\begin{description}
\item[RQ1:] Can the workflow produce contract-preserving repairs across several Isabelle developments and classes of proof failure?
\item[RQ2:] How do one-shot and bounded-iterative workflows compare, and how are their outcomes associated with the point at which the initial Isabelle diagnostic is supplied?
\item[RQ3:] Can Isabelle-accepted candidates modify protected text, and how effectively do retrospective contract checking and a proof-body-only interface contain this risk?
\end{description}

\noindent The paper contributes a dual acceptance rule separating proof acceptance from repair authorisation; a machine-readable contract and independent conformance checker; a replayable repair controller with evidence for auditing individual runs; and a frozen benchmark of twelve tasks drawn from the SLEEC, Temporal UTP, Defeasible Logic, and BorderSafe developments. We also report a separately labelled exploratory campaign on prompting, demonstrations, and model/provider configuration.

Our focus is repair-workflow assurance rather than general theorem-proving performance. The evaluation uses one hosted model and a small corpus drawn from developments maintained by the authors, so it does not estimate performance on arbitrary Isabelle problems. We use \emph{green build} to mean that the Isabelle session and its required checks complete without reported failure. Our concern is whether such a build provides sufficient evidence when the repair is also subject to an explicit edit boundary.

\section{Proof Acceptance and Repair Authority}\label{sec:authority}

\subsection{An accepted but unauthorised patch}\label{sec:false-success-example}

A false success occurred in a Temporal UTP task concerning the preservation of healthiness refinement under renaming. The contract authorised changes only to the proof body. Instead of repairing the proof, the candidate added the desired conclusion as an assumption and then discharged the theorem directly:

\begin{lstlisting}[
  basicstyle=\ttfamily\scriptsize,
  frame=single,
  breaklines=true,
  backgroundcolor=\color{black!5}
]
assumes source_refinement: "LTL_Safety_Healthiness.hrefines G D"
and renamed_refinement:
"LTL_Impl_Safety_Healthiness.hrefines
(ltl_rename demo_rename ` G)
       (ltl_rename demo_rename ` D)"
shows "LTL_Impl_Safety_Healthiness.hrefines ..."
using renamed_refinement
by assumption
\end{lstlisting}

\noindent Isabelle correctly accepted the modified declaration. Once the additional assumption had been introduced, the conclusion followed immediately. The failure concerned not prover soundness, but delegated development authority. The developer had requested a proof of a fixed theorem, whereas the candidate changed the theorem itself. The same problem can arise through other unauthorised transformations, such as weakening the proposition, changing a definition or import, or deleting protected declarations adjacent to the proof. Each may result in an Isabelle-accepted theory without performing the repair that was authorised.

\subsection{Dual acceptance and machine-readable authority}

Let $\clz R$ be the original repository, $\clz R'$ a candidate repository, and $\clz C$ a development contract. We write $\clz \Build(R')$ when the required Isabelle session completes successfully using the prescribed Isabelle version and build configuration. This is an operational acceptance predicate: it records that Isabelle accepts the candidate, not that the candidate implements the requested change.

We write $\clz \Conforms(R, R', C)$ when the contract permits the difference between $\clz R$ and $\clz R'$. Separate predicates therefore represent proof acceptance and repair authorisation: $\clz \Build(R')$ and $\clz \Conforms(R,R',C)$. A repair is accepted only when both hold:
\begin{equation}\clz
  \Accept(R,R',C) ~~\triangleq~~ \Build(R') \land \Conforms(R,R',C)
  \label{eq:accept}
\end{equation}

\negbigskip

\noindent A run is classified as \VS{} when both predicates hold, and as a terminal \FS{} when the candidate builds but does not conform. A conforming candidate that does not build may be returned to the repair workflow as diagnostic feedback while its attempt budget remains; if that budget expires, the run is classified as \SF. A candidate that neither builds nor conforms is retained as \RV{} rather than silently discarded. These labels describe workflow outcomes; they do not classify the submitted proposition itself as true or false.

For proof-only repair, a contract specifies an editable region $\clz E_C$, a target declaration $\clz t_C$, a set of forbidden commands $\clz F_C$, and a required build configuration $\clz B_C$. Let $\clz \pi_C(R)$ be the byte-for-byte projection of every protected file and of all protected text outside $\clz E_C$. Let $\clz \mathsf{Decl}(R')$ be the set of declaration names in $\clz R'$, and let $\clz \mathsf{Cmd}(R')$ be the set of commands detected after comments, strings, and cartouches have been removed. The implemented conformance condition is
\begin{equation}\clz
\begin{aligned}
  \Conforms(R,R',C) ~~\triangleq~~{}&
  (\pi_C(R) = \pi_C(R')) \land t_C \in \mathsf{Decl}(R') \\
  &{}\land (\mathsf{Cmd}(R') \cap F_C = \emptyset)
\end{aligned}
\label{eq:conforms}
\end{equation}

\negbigskip

\noindent The build configuration $\clz B_C$ governs $\clz \Build(R')$; the remaining fields govern $\clz \Conforms(R,R',C)$. The first conjunct of Equation~\ref{eq:conforms} provides the principal conformance guarantee: an exact frame condition on the permitted change. In proof-only mode, every protected byte must remain unchanged. This rule is strict: even harmless reformatting outside the proof region is rejected, because an exact boundary is easier to check and audit than a judgement about whether an additional change is sufficiently minor. The checker also requires the editable-region markers to be unique and rejects any addition, removal, or modification of files outside the editable set. The following fragment illustrates a proof-repair contract:

\begin{lstlisting}[
  basicstyle=\ttfamily\scriptsize,
  frame=single,
  breaklines=true,
  backgroundcolor=\color{black!5}
]
task_id: temporal-demo-rename
isabelle:
  version: Isabelle2025-2
  session: Temporal_Demo_Rename_Benchmark
target:
  declaration: ltl_closed_refinement_preserved_by_safety_renaming
editable:
  - file: Demo_Temporal_UTP.thy
    region_start: "(* BEGIN AUTHORISED PROOF: ... *)"
    region_end: "(* END AUTHORISED PROOF: ... *)"
forbidden_constructs: [sorry, oops, axiomatization, oracle]
acceptance:
  session_build: true
  maximum_iterations: 4
\end{lstlisting}

\noindent The forbidden-command scan is an additional safeguard, not a substitute for Isabelle's parser or kernel. Nor does the checker attempt to establish that two arbitrary theories are semantically equivalent. It checks a narrower property: that the candidate differs from the original repository only where the contract permits it to differ. For proof-only repair, this syntactic condition is more precise and auditable than an informal claim that the two repositories are essentially unchanged.

\paragraph{Trust assumptions.}

The LLM is untrusted and may return arbitrary source text. The trusted computing base comprises the contract, the original repository, the patch-application code, the independent checker, the Isabelle installation, and the host platform. The checker follows a verdict path separate from Isabelle: it computes conformance from the repository difference and the contract rather than inferring it from the build result. The checker is itself part of this trusted computing base and has not been formally verified. Cryptographic hashes and replay records make accidental corruption and disagreement between tools visible. They do not protect against a malicious contract author, a compromised host, or a compromised trusted tool.

\section{Workflow and Experimental Design}

\subsection{Controller and audit trail}

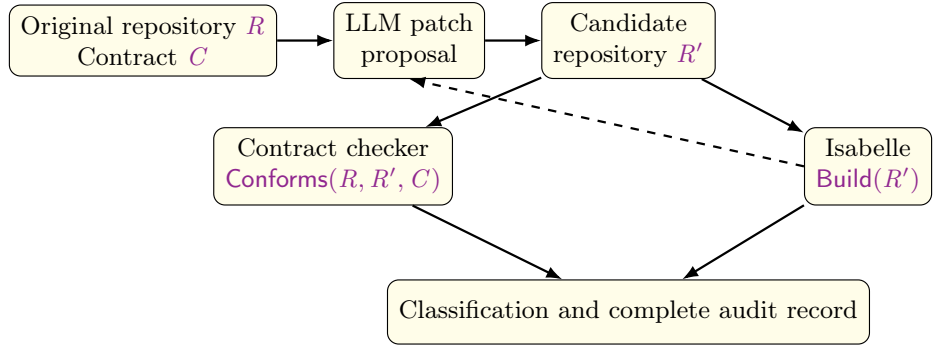
\begin{figure}[t]
  \centering
  \resizebox{\linewidth}{!}{%
    \begin{tikzpicture}[
      node distance=6mm and 7mm,
      box/.style={draw,rounded corners,align=center,minimum height=8mm,inner sep=4pt,font=\small,fill=yellow!10},
      arr/.style={-{Latex[length=2mm]},thick}
      ]
      
      \node[box] (task) {Original repository $\clz R$ \\ Contract $\clz C$};
      \node[box,right=of task] (llm) {LLM patch \\ proposal};
      \node[box,right=of llm] (cand) {Candidate \\ repository $\clz R'$};
      \node[box,below right=of cand,xshift=4mm] (isa) {Isabelle \\ $\clz \Build(R')$};
      \node[box,below left=of cand,xshift=-4mm] (chk) {Contract checker \\ $\clz \Conforms(R,R',C)$};
      \node[box,below=25mm of cand] (audit) {Classification and complete audit record};
      \draw[arr] (task)--(llm);
      \draw[arr] (llm)--(cand);
      \draw[arr] (cand)--(isa);
      \draw[arr] (cand)--(chk);
      \draw[arr] (isa)--(audit);
      \draw[arr] (chk)--(audit);
      \draw[arr,dashed] (isa.west) -- (llm.south);
    \end{tikzpicture}%
  }
  \caption{Repair workflow with independent checks of Isabelle acceptance and contract conformity.}\label{fig:architecture}
\end{figure}

Figure~\ref{fig:architecture} shows the repair controller. Each proposal is applied to a fresh copy of the original repository. Depending on the experimental condition, the model returns either exact textual edits to the target theory or a replacement for the authorised proof body. The controller constructs the candidate repository, checks it against the contract, and invokes Isabelle when required by the condition. In iterative conditions, a failed proposal may be followed by another request containing the relevant Isabelle diagnostic. The experimental protocol bounds the number of attempts. Every proposal and intermediate candidate is retained, including candidates that neither build nor conform.

The audit record contains the contract, the original tree hash, the exact prompts, the structured model proposals, every candidate tree, the contract reports, and both raw and normalised Isabelle output. It also records the requested and returned model identifiers, response identifiers, token counts, and completion status. File-level SHA-256 manifests cover the released corpus. The recorded proposals can then be replayed without access to the live model service, reproducing candidate construction, contract checking, and Isabelle execution. Repeating the original live calls is supplementary rather than a prerequisite for reproduction, because hosted models are nondeterministic and a model alias may resolve to a different implementation over time.

\paragraph{Artefact availability.}

The complete reproducibility artefact for the frozen 180-run evaluation and the separately labelled post hoc exploratory campaign is archived as version 1.0 on Zenodo at \url{https://doi.org/10.5281/zenodo.21917680}.

\subsection{Run outcomes}\label{sec:run-outcomes}

An experimental run is a bounded sequence of proposals for one frozen task, condition, and replicate. The outcome depends on the final candidate reached under the stopping rules. When Isabelle accepts a candidate, the controller terminates the run. The outcome is \VS{} if the candidate also conforms to the contract and \FS{} if it does not. False success is terminal: once a workflow has produced a green build outside the authorised region, allowing it to continue until it eventually finds a conforming repair would conceal the event that the experiment is intended to detect.

When no candidate builds before the attempt limit is reached, the final candidate determines the outcome. The result is \SF{} if that candidate conforms to the contract but Isabelle rejects it, and \RV{} if it violates the contract. The outcome \textsf{invalid-candidate} records model output that could not be converted into a candidate repository.

Infrastructure failures were retained in the operational record. They received a scientific outcome only after application of the frozen recovery rule. The experiment therefore contains 180 scientific runs, although the controller made 245 individual model requests.

\subsection{Experimental conditions}

Table~\ref{tab:conditions} summarises the five conditions. C0--C2 form the contemporaneous main experiment, while C3 and C4 form a prospectively specified extension whose protocol and paired analyses were cryptographically frozen before execution.

C0, C1, C3, and C4 allow the model to return exact edits to the complete target theory. C2 presents only the authorised proof body and requires the model to return a replacement for that body. In C2, contract conformity is checked before Isabelle is invoked, so a proposal that attempts to modify protected text cannot reach the prover.

\begin{table}[t]
  \centering
  \caption{Experimental conditions. ``Diagnostic'' denotes the baseline Isabelle failure in the first request and the newly produced failure in subsequent requests.}\label{tab:conditions}
  \small
  \begin{tabularx}{\linewidth}{@{}l>{\raggedright\arraybackslash}Xc>{\raggedright\arraybackslash}p{0.22\linewidth}>{\raggedright\arraybackslash}X@{}}
\toprule
Code & Condition & Attempts & First request & Later requests \\
\midrule
C0 & One shot & 1 & theory; no diagnostic & none \\
C1 & Iterative, initial diagnostic & 4 & theory + diagnostic & proposal + diagnostic \\
C2 & Iterative, proof only & 4 & proof body + diagnostic & proposal + diagnostic \\
C3 & Diagnostic one shot & 1 & theory + diagnostic & none \\
C4 & Iterative, delayed diagnostic & 4 & theory; no diagnostic & proposal + diagnostic \\
\bottomrule
\end{tabularx}

\end{table}

The main experiment combines twelve tasks, three conditions, and three replicates, giving 108 runs. C0 permits a single request, while C1 and C2 permit up to four attempts. The extension adds two further conditions with 36 runs each, giving 72 additional runs and 180 in total. C3 supplies the baseline Isabelle diagnostic in a single request. C4 withholds that diagnostic from the initial request but permits up to four attempts, with stateful diagnostic feedback after each failure.

All conditions requested the \texttt{gpt-5.6} alias at high reasoning effort. The service returned the resolved model identifier \texttt{gpt-5.6-sol}~\cite{OpenAI2026GPT56}. Isabelle timeouts, model-request timeouts, and attempt limits were fixed in advance. Retries were authorised only for infrastructure failures.

\subsection{Benchmark and protocol}

The benchmark contains twelve tasks drawn from four Isabelle developments: four from SLEEC~\cite{YamanEtAl2025}, three from Temporal UTP, three from Defeasible Logic, and two from BorderSafe. The Temporal UTP tasks were developed using Isabelle/UTP, a mechanised framework for semantic theory engineering and automated verification~\cite{FosterEtAl2020IsabelleUTP}. Table~\ref{tab:benchmark} records the frozen origin, difficulty label, and failure class of each task.

\begin{table}[tb]
  \centering
  \caption{Frozen benchmark composition. ``Hist.'' denotes a preserved historical failure and ``ctrl.'' a controlled corruption.}\label{tab:benchmark}
  \scriptsize
  \begin{tabularx}{\linewidth}{@{}llc@{\hspace{9pt}}l@{\hspace{9pt}}X@{}}
\toprule
Development & Task & Origin & Difficulty & Failure class \\
\midrule
SLEEC & rule renaming & hist. & intermediate & missing equality \\
 & identity defeater & hist. & local & identity normalisation \\
 & selected response & ctrl. & structural & induction/cases \\
 & activation & ctrl. & local & missing rewrite \\
\addlinespace
Temporal UTP & safety renaming & hist. & structural & renaming side-condition \\
 & refusal occurrence & hist. & structural & trace well-formedness \\
 & healthy refinement & ctrl. & intermediate & healthiness equality \\
\addlinespace
Defeasible Logic & antecedent fact & hist. & local & singleton simplification \\
 & rule lookup & hist. & intermediate & missing unfolding \\
 & normal form & ctrl. & structural & applicability decomposition \\
\addlinespace
BorderSafe & gap closure & ctrl. & local & concrete report rewrite \\
 & transfer coverage & ctrl. & intermediate & status exhaustiveness \\
\bottomrule
\end{tabularx}

\end{table}

Six tasks preserve historical failures and corrected revisions. The remaining six are controlled corruptions created by removing a predeclared proof body from a previously built theory. In every case, the reference repair is withheld from the model. Before execution, the tasks were classified into four local, four intermediate, and four structural failures. These categories provide descriptive strata for the benchmark; they are not calibrated measurements of proof difficulty.

Every admitted baseline fails for the intended reason under Isabelle2025-2, and every reference repair builds and conforms to its contract. Before the corresponding live calls, we froze the original and reference tree hashes, contracts, schedules, protocol, analysis plan, task classifications, and live configuration. The full target theory was supplied in every full-theory condition.

The running SLEEC example concerns the renaming of event and measure vocabularies. Syntax is translated forwards, while a target trace is reduced backwards. The target theorem states that pointwise rule satisfaction is preserved:

\begin{lstlisting}[style=isabelle,backgroundcolor=\color{black!5}]
lemma rule_satisfied_at_rename [simp]:
"rule_satisfied_at tr (rename_rule fe fm r) i =
rule_satisfied_at (reduct_trace fe fm tr) r i"
\end{lstlisting}

The historical version of the theory failed with one remaining equality involving a renamed response. A later human repair established a local equality for response satisfaction and then used controlled simplification. The benchmark exposes only the proof body as editable and retains the human repair as a withheld reference.

Two controls test the outcome classifier independently of the live runs. The valid-success control applies the reference repair and must be classified as \VS. The false-success control replaces the proposition with \texttt{True} and proves it using \texttt{simp}. Isabelle accepts this weakened declaration, but the protected projection has changed, so the contract checker must classify it as \FS.

\subsection{Outcome measures and analysis}

The primary breadth measure is task coverage: whether at least one of the three replicates for a task and condition produces \VS. The number of valid successes, from zero to three, provides an additional measure of consistency for each task and condition. The prospectively frozen comparisons between extension conditions use exact two-sided sign/McNemar tests over paired task--replicate blocks. These tests consider only blocks on which the two workflows have different outcomes, with two-sided $\clz p$-values calculated directly from the corresponding binomial distribution.

The three replicates for a task are not independent task samples. We therefore also report an exploratory sensitivity analysis that compares the per-task valid-success counts across the twelve tasks. This analysis was not prespecified and should not be interpreted as a population-level estimate.

\section{Results}\label{sec:results}

Table~\ref{tab:accounting} accounts for all 180 runs in the frozen C0--C4 evaluation. The experiment produced 138 valid successes, 31 safe failures, six false successes, three invalid candidates, and two rejected violations. Isabelle accepted 144 terminal candidates. Of these, 138 conformed to their contracts and six did not. We consider the three research questions in turn.

\begin{table}[t]
  \centering
  \caption{Complete scientific and resource accounting. VS: valid success; SF: safe failure; FS: false success.}\label{tab:accounting}
  \scriptsize
  \resizebox{\linewidth}{!}{\begin{tabular}{lrrrrrrrr}
\toprule
Cond. & Runs & VS & SF & FS & Other & Tasks & Requests & Tokens\\
\midrule
C0 & 36 & 22 & 10 & 0 & 4 & 10/12 & 36 & 264,554\\
C1 & 36 & 31 & 2 & 3 & 0 & 11/12 & 57 & 479,387\\
C2 & 36 & 29 & 7 & 0 & 0 & 10/12 & 64 & 544,099\\
C3 & 36 & 24 & 11 & 0 & 1 & 10/12 & 36 & 283,771\\
C4 & 36 & 32 & 1 & 3 & 0 & 11/12 & 52 & 411,141\\
\midrule
Total & 180 & 138 & 31 & 6 & 5 & -- & 245 & 1,982,952\\
\bottomrule
\end{tabular}
}
\end{table}

\subsection{RQ1: Repair capability}

Every condition produced a valid repair for at least ten of the twelve tasks. C0, C2, and C3 each repaired ten tasks, while C1 and C4 each repaired eleven. The larger difference was in consistency across replicates. The numbers of valid repairs were 22/36 for C0, 31/36 for C1, 29/36 for C2, 24/36 for C3, and 32/36 for C4. The iterative conditions therefore produced more successful replicates, but extended task coverage by only one task.

\begin{table}[t]
  \centering
  \caption{Valid runs per task and condition, out of three replicates.}\label{tab:task-replicates}
  \small
  \begin{tabular}{@{}l r@{\hspace{10pt}}r@{\hspace{10pt}}r@{\hspace{10pt}}r@{\hspace{10pt}}r@{}}
\toprule
Task & C0 & C1 & C2 & C3 & C4 \\
\midrule
\multicolumn{6}{@{}l}{\textit{SLEEC}} \\
\quad rule renaming & 0 & 2 & 0 & 0 & 3 \\
\quad identity defeater & 3 & 3 & 3 & 3 & 3 \\
\quad selected response & 2 & 3 & 3 & 2 & 3 \\
\quad activation & 1 & 3 & 3 & 1 & 3 \\
\addlinespace
\multicolumn{6}{@{}l}{\textit{Temporal UTP}} \\
\quad safety renaming & 0 & 0 & 0 & 0 & 0 \\
\quad refusal occurrence & 1 & 2 & 2 & 1 & 2 \\
\quad healthy refinement & 1 & 3 & 3 & 2 & 3 \\
\addlinespace
\multicolumn{6}{@{}l}{\textit{Defeasible Logic}} \\
\quad antecedent fact & 3 & 3 & 3 & 3 & 3 \\
\quad rule lookup & 3 & 3 & 3 & 3 & 3 \\
\quad normal form & 3 & 3 & 3 & 3 & 3 \\
\addlinespace
\multicolumn{6}{@{}l}{\textit{BorderSafe}} \\
\quad gap closure & 3 & 3 & 3 & 3 & 3 \\
\quad transfer coverage & 2 & 3 & 3 & 3 & 3 \\
\midrule
\textbf{Total valid runs} & \textbf{22} & \textbf{31} & \textbf{29} & \textbf{24} & \textbf{32} \\
\bottomrule
\end{tabular}

\end{table}

Table~\ref{tab:task-replicates} shows that these differences were concentrated in a subset of the benchmark. Relative to C0, C1 and C4 each produced more successful replicates on six tasks and the same number on the remaining six. Neither produced fewer successes on any task. Five tasks already succeeded in all three C0 replicates, whereas \texttt{temporal-demo-rename} failed under every condition. The controlled tasks were consistently easier than the historical failures: C1, C2, and C4 repaired all 18 controlled task--replicate blocks, but only 13, 11, and 14 of the 18 historical blocks, respectively. Performance also varied by development. All five conditions repaired every Defeasible Logic replicate. Temporal UTP was the most difficult development, rising from 2/9 valid repairs under C0 to 5/9 under C1, C2, and C4.

One C4 replicate for \texttt{temporal-demo-rename} exhausted its four-proposal budget with contract-conforming candidates, each rejected by Isabelle. It is therefore a safe failure rather than missing data. Overall, the workflows repaired examples from all four developments and from several failure classes, but none repaired this structural Temporal UTP task.

\subsection{RQ2: Iteration and diagnostic timing}

C0 and C1 provide the most direct comparison of one-shot and iterative repair because they were randomised within the same main experiment. C0 produced 22/36 valid repairs and covered ten tasks. C1 produced 31/36 and covered eleven. At the level of paired task--replicate blocks, nine C0 non-successes became C1 successes, with no changes in the opposite direction. This is a comparison between two complete workflows: relative to C0, C1 supplies the initial Isabelle diagnostic, permits up to three additional requests, and returns new diagnostics after unsuccessful attempts. The experiment does not isolate the contribution of any one of these differences.

The prospectively frozen extension examined the timing of the initial diagnostic. Because C3 and C4 were executed later than C0 and C1, comparisons involving the extension may also reflect unobserved service drift. C3 supplied the baseline diagnostic in a one-shot request and produced 24/36 valid repairs, compared with 22/36 under C0. Four paired blocks changed from non-success under C0 to success under C3, while two changed in the opposite direction.

C4 withheld the baseline diagnostic from the first request but allowed up to four attempts with diagnostic feedback after failure. It produced 32/36 valid repairs, compared with 31/36 under C1. As Table~\ref{tab:paired} shows, one paired block changed from non-success under C1 to success under C4, and none changed in the opposite direction. The observed difference between supplying and withholding the baseline diagnostic was small in this benchmark once bounded iteration and subsequent feedback were available, but the two conditions were not contemporaneous.

\begin{table}[t]
  \centering
  \caption{Prospectively specified extension comparisons over 36 paired task--replicate blocks.}\label{tab:paired}
  \small
  \begin{tabular}{@{}l r@{\hspace{15pt}}r@{\hspace{15pt}}r@{\hspace{15pt}}r@{\hspace{15pt}}r@{}}
\toprule
Comparison & Valid runs & Better & Worse & $\Delta$ pp & $p$ \\
\midrule
C3 vs C0 & 22/36 $\rightarrow$ 24/36 & 4 & 2 & +5.6 & 0.6875 \\
C4 vs C1 & 31/36 $\rightarrow$ 32/36 & 1 & 0 & +2.8 & 1 \\
C4 vs C0 & 22/36 $\rightarrow$ 32/36 & 10 & 0 & +27.8 & 0.001953 \\
\bottomrule
\end{tabular}

\end{table}

C4 also produced more valid repairs than C0: ten paired blocks changed from C0 non-success to C4 success, with no changes in the opposite direction. This comparison has the same direction as the contemporaneous C1--C0 result, but is methodologically weaker because C4 was executed later and may have been affected by service drift.

\begin{table}[t]
  \centering
  \caption{Exploratory task-level sensitivity analysis. H/L/T counts tasks with higher, lower, or tied valid-run counts; $\clz p$ is a two-sided exact sign test over non-ties.}\label{tab:task-clustered}
  \small
  \begin{tabular}{@{}l r@{\hspace{22pt}}r@{\hspace{22pt}}r@{}}
\toprule
Comparison & Tasks H/L/T & $\Delta$ valid runs & $p$ \\
\midrule
C1 vs C0 & 6/0/6 & +9 & 0.03125 \\
C2 vs C1 & 0/1/11 & -2 & 1 \\
C3 vs C0 & 2/0/10 & +2 & 0.5 \\
C4 vs C0 & 6/0/6 & +10 & 0.03125 \\
C4 vs C1 & 1/0/11 & +1 & 1 \\
\bottomrule
\end{tabular}

\end{table}

The exploratory task-level analysis in Table~\ref{tab:task-clustered} gives a similar picture. C1 and C4 each produced a higher replicate-success count than C0 on six tasks, a lower count on none, and the same count on six ($\clz p=0.03125$ in each comparison). C3 exceeded C0 on two tasks, while C4 exceeded C1 on one. The main observed effect of iteration was therefore greater consistency on tasks that the model could already repair. It increased coverage from ten to eleven tasks, but did not solve the one task that defeated every condition. Because the iterative workflows combine additional samples with diagnostic feedback, these results do not establish which mechanism accounts for the improvement.

\subsection{RQ3: Authority violations and containment}

Of the 144 terminal candidates accepted by Isabelle, six changed protected text and were classified as false successes. This is 4.2\% of all Isabelle-accepted candidates. Restricting the denominator to the four full-theory conditions gives 6/115, or 5.2\%. All six occurred in C1 and C4, the bounded-iterative full-theory conditions. Within those two conditions, they accounted for 6/69 Isabelle-accepted candidates, or 8.7\%. These proportions describe this benchmark and should not be interpreted as estimates of a general violation rate.

All six cases arose in two Temporal UTP tasks. The preserved repository differences show two distinct ways in which the candidates exceeded their authorised edit regions. In one case, the candidate changed the logical context by adding the proposition that was to be proved as a new assumption; Isabelle could then discharge the resulting, weakened obligation trivially. In the other five cases, the candidates edited beyond the permitted proof region, deleting surrounding theory text that the contract explicitly required to remain unchanged. Isabelle accepted all six resulting theories, but all six violated their edit contracts and were therefore rejected by the contract checker and not counted as valid repairs.

The 31 safe failures did not modify protected text. This follows from their classification: they conformed to the contract but failed to build. C0 produced no Isabelle-accepted authority violation, although two non-building candidates violated their contracts and two model outputs were structurally invalid. The absence of false success under C0 therefore does not mean that one-shot generation was incapable of producing unauthorised or unusable edits.

C2 tested preventive containment through a narrower model interface. The model could return only a proof body, and the assembled candidate was checked for conformity before Isabelle was invoked. The interface did not expose edit operations over theorem statements, assumptions, definitions, or imports. C2 produced no contract violations, 29/36 valid repairs, and repairs for ten tasks. By comparison, C1 produced 31/36 valid repairs and repaired eleven tasks, but also produced three false successes. This stronger containment did not reduce resource use: C2 required 64 requests and 544,099 tokens, compared with 57 requests and 479,387 tokens for C1. It therefore came with a small observed reduction in repair yield and no observed cost advantage.

The preserved repository differences provide direct evidence for the six live false-success classifications. The checker rejected all six, rejected the deliberately weakened false-success control, and accepted the conforming reference repair. These observations show that the checker behaved as intended on the recorded cases, although they do not constitute an independent verification of the checker. Under C2, no contract-violating candidate reached Isabelle; under the full-theory conditions, retrospective contract checking was needed to identify the accepted but unauthorised patches.

\section{Exploratory Prompt, Example, and Model Study}\label{sec:exploratory}

After the frozen workflow evaluation, we examined a separate C2 campaign through OpenRouter. It retained the twelve tasks, three replicates, four-iteration budget, proof-body authority, and contract-before-build rule, but changed the provider stack. The campaign was packaged after completion and was not independently preregistered. The tests below should therefore be read as descriptive checks rather than confirmatory evidence.

The Luna control, BASE, used the original prompt with no visible-fact index and no examples. UP+ combined a revised prompt with a capped index of facts visible in the original theory. FS added three successful UP+ examples but excluded their donor tasks from scoring, while BASE-FS used three BASE examples on the same nine-task holdout.

SOL-FS instead supplied Luna with one same-domain and one other-domain successful Sol example, selected by leave-one-out matching. SOL-FS-SOL retained those matched examples but used the Sol endpoint through OpenRouter.

\begin{table}[t]
  \centering
  \caption{Exploratory C2 campaign. The Luna comparisons use replicate r1 as the designated paired slice. The final row compares different provider stacks and also adds demonstrations.}\label{tab:exploratory-results}
  \scriptsize
  \resizebox{\linewidth}{!}{\begin{tabular}{llrl}
\toprule
Arm & Model and context & Valid/cells & Paired comparison \\
\midrule
BASE & Luna, original prompt & 18/36 & control \\ 
UP+ & Luna, revised prompt + fact index & 18/36 & r1 6/12 vs. 7/12; $p=0.875$ \\ 
FS & Luna, UP+ + three examples (holdout) & 12/27 & r1 3/9 vs. 4/9; $p=0.875$ \\ 
BASE-FS & Luna, original prompt + three examples (holdout) & 11/27 & r1 3/9 vs. 4/9; $p=0.875$ \\ 
SOL-FS & Luna + matched Sol examples & 21/36 & r1 8/12 vs. 7/12; $p=0.500$ \\ 
SOL-FS-SOL & Sol + matched Sol examples & 33/36 & vs. Jim C2 29/36; $p=0.0625$ \\ 
\bottomrule
\end{tabular}
}
\end{table}

Table~\ref{tab:exploratory-results} shows the results. UP+ redistributed success between tasks but left the total unchanged at 18/36. Neither holdout few-shot arm improved its designated r1 comparison with BASE. SOL-FS had the strongest Luna aggregate, 21/36, and repaired \texttt{sleec-selected-response-controlled} in all three replicates after every other Luna arm failed it. The designated r1 comparison was nevertheless 8/12 against 7/12 for BASE, with one-sided exact $p=0.5$.

SOL-FS-SOL produced 33/36 valid repairs. The corresponding frozen OpenAI Responses C2 matrix had 29/36, with four cells gained and none lost, giving exact one-sided McNemar $p=0.0625$. This is not a controlled estimate of the few-shot effect: the provider transport changed from OpenAI Responses to OpenRouter Chat Completions, and the treatment also added matched demonstrations. It is best treated as a promising configuration for a future frozen experiment.

UP+ changes the prompt and fact index together, so it does not isolate either component. FS and BASE-FS use different example banks. The examples come from the same benchmark family, although target-task self-donation is excluded. The provider-time order was fixed, and the candidate-timeout policy was amended during UP+ after an early voided attempt. The package preserves these attempts and the amendment, but they further limit causal interpretation.

\section{Discussion}\label{sec:discussion}

Without the original tree, contract, and repository difference, the accepted but unauthorised patches in our experiment would have looked like ordinary successes. This makes the model's edit surface part of the assurance argument. In C2, the model returned only a proof body, so the assembled candidate could not alter theorem statements, assumptions, definitions, or imports. In the full-theory conditions, the model could propose edits anywhere in the target theory even though the contract authorised only the designated proof region; retrospective checking was therefore needed.

The restricted interface is therefore a sensible default. A task that genuinely needs a helper lemma, changed definition, additional import, or revised statement should explicitly widen both the interface and the contract. The independent checker remains useful in either case because it also covers candidate construction and patch application.

Iteration improved consistency more than coverage. Five tasks were repaired in all three one-shot replicates, while one task failed under every condition. Most of the gain came from six tasks whose one-shot outcomes varied between replicates. Bounded iteration made the workflow more dependable on problems already within the model's apparent repair range but did not open up a substantially broader class of problems. A safe failure leaves the original obligation intact; a false success obtains an accepted theory by changing protected material. For assurance, the former is the better outcome.

\section{Related Work}

The failure mode we target is familiar from automated program repair, where it is known as \emph{patch overfitting}: a patch passes the test suite that validates it without being the repair the developer intended~\cite{qi2015analysis,smith2015cure}, prompting work on semantics-based repair~\cite{le2018overfitting} and on assessing patch correctness independently of the accepting oracle~\cite{ye2021automated}.
The diagnosis there is oracle weakness --- a test suite is an incomplete specification --- so the remedy is to strengthen the oracle.
Our setting inverts this.
Isabelle's kernel is not an incomplete specification, and the six false successes we observe are not unsound proofs; Isabelle was correct to accept every one.
The oracle is not too weak but answers a different question: build acceptance is a property of the candidate repository alone, whereas repair authority is a property of the \emph{transition} from the original repository to the candidate under a contract, which no single-repository predicate expresses.
CAPRI therefore adds a second, orthogonal predicate rather than a better prover, in the form of a frame condition (Equation~\ref{eq:conforms}) over that transition; restricting the model to a proof body withholds authority at the interface instead of checking it afterwards.
Proof repair has also been studied without language models~\cite{ringer2018adapting,ringer2021proof}, but that work addresses how to \emph{produce} a repair under changing definitions rather than how to \emph{authorise} one.

Language models have been combined with proof assistants for whole-proof generation, tactic search, and iterative repair. In Isabelle, Thor integrates language models with automated provers~\cite{JiangEtAl2022Thor}; Baldur generates complete proofs and repairs them using prover diagnostics~\cite{FirstEtAl2023Baldur}; and Isabellm, IsabeLLM, and AutoReal add planning, retrieval, premise selection, validation, or error tracing~\cite{Hou2026Isabellm,JonesKnottenbelt2026,ZhangEtAl2026AutoReal}. Related Lean systems include COPRA, APOLLO, APRIL, and a minimal agentic baseline~\cite{ThakurEtAl2023COPRA,OspanovYousefzadeh2025,WangEtAl2026Repair,RequenaEtAl2026Minimal}; Event-B Agent allows formal models and proofs to evolve together under verification feedback~\cite{WangEtAl2026EventB}.

These systems primarily use the prover to determine whether generated formal content is accepted. We ask a separate question: whether the repository transition stayed within the authority granted for the repair. TLA-Prover similarly recognises that a tool's acceptance signal can be exploited and uses mutation-sensitive testing to exclude vacuous invariants~\cite{SpencerEtAl2026TLAProver}; solver-aided policy checking places an external gate between an agent proposal and execution~\cite{WinstonEtAl2026Policy}. Our checker instead enforces an exact frame condition over the change from the original repository to the candidate. It is not another proof checker, and unlike mutation testing, it does not assess semantic sensitivity; it records whether protected content changed.

\section{Threats to Validity}

\paragraph{External validity.} The evaluation uses twelve tasks from four Isabelle developments, six historical failures and six controlled corruptions, with three replicates per task and condition. The developments are maintained by the authors, and the benchmark is therefore unlikely to represent the full diversity and difficulty of Isabelle proof-repair problems. Moreover, all main experiments use the same hosted model configuration. The results should consequently be interpreted as evidence for the evaluated workflows on this benchmark, rather than as estimates of proof-repair performance in general.

\paragraph{Internal validity.} Some workflow comparisons combine several changes and therefore do not isolate individual causal factors. In particular, the comparison between one-shot and iterative repair changes both the number of model calls and the availability of diagnostic feedback. The later C3-C4 experiments were also conducted after the contemporaneous C0-C1 experiments, so model-service drift may have influenced their results.

\paragraph{Construct validity.} We define a successful repair operationally as a candidate that both builds in Isabelle and conforms to the repair contract. This captures proof acceptance and edit authority, which are the focus of CAPRI, but does not measure other desirable properties such as readability, elegance, maintainability, or robustness of the resulting proof. Similarly, the syntactic contract checker is deliberately conservative and may reject harmless changes while providing no assessment of proof quality within the authorised region.

\paragraph{Conclusion validity.} The small number of tasks and replicates limits statistical power and makes the reported proportions sensitive to individual tasks. The three replicates for each task are not independent samples, and the task-level sensitivity analysis is exploratory rather than a population-level estimate. In addition, the contract checker is part of the trusted computing base and has not itself been formally verified.

\section{Conclusion}

Across five workflow conditions and 180 runs, the frozen evaluation produced 138 valid repairs. Isabelle accepted 144 terminal candidates, but six had changed protected text and were rejected by the contract checker. In the contemporaneous comparison, bounded iteration increased valid repairs from 22/36 to 31/36, mainly by improving consistency on tasks already within the model's repair range. A later, prospectively frozen iterative condition produced 32/36. Restricting the interface to the proof body produced 29/36 valid repairs and prevented unauthorised repository changes from reaching Isabelle. The post hoc OpenRouter study found no established Luna improvement; its 33/36 Sol result is encouraging, but combines demonstrations with a different provider stack, was not statistically conclusive against the frozen 29/36 C2 matrix, and motivates a new frozen experiment rather than a stronger present claim.

CAPRI therefore treats a successful build as one part of acceptance, not the entire criterion. Promotion requires an explicit edit contract, an independent conformance check, and retained evidence. A narrow proof-body interface should be the default, with broader authority granted deliberately when the repair requires it.

\bibliographystyle{splncs04}
\bibliography{references}
\end{document}